\documentclass{article}

\newcommand{\bear}{\begin{eqnarray}}
\newcommand{\eear}{\end{eqnarray}}
\newcommand{\be}{\begin{equation}}
\newcommand{\ee}{\end{equation}}
\newcommand{\beqn}{\begin{eqnarray}}
\newcommand{\eeqn}{\end{eqnarray}}
\newcommand{\beqnn}{\begin{eqnarray*}}
\newcommand{\eeqnn}{\end{eqnarray*}}

\def\vep{\varepsilon}
\def\vf{\varphi}

\usepackage{cite}
 
\begin{document}

\begin{center} { \bf
Coherent states in a magnetic field and their generalizations } \end{center}

\begin{center} {\bf
V. V. Dodonov 
}\end{center}


\begin{center}
{\it
Institute of Physics and International Center for Physics, University of Brasilia,
 Brazil }

E-mail:~~~vdodonov@fis.unb.br\\
\end{center}

\abstract{
This is a brief review of various families of coherent and squeezed states
(and their generalizations)
for a charged particle in a magnetic field, that have been constructed 
for the past 50 years.
Although the main attention is paid to the Gaussian states, various families of 
non-Gaussian states are also discussed, and the list of relevant references is provided.}

\section
{Introduction}

Superposition states (wave packets) of charged particles moving in a magnetic field attracted attention 
of many researchers for many reasons. First, some of them can be considered as the simplest 
non-trivial two-dimensional generalizations of the coherent states of a harmonic
oscillator of the Schr\"odinger--Klauder--Glauber--Sudarshan type.
Second, there exist many different families of coherent and other superposition states, originating from
the infinite degeneracy of the energy spectrum in the absence of a confining potential.
Also, such states are interesting from the point of view of quantum mechanics on the 
{\em non-commutative plane}. The goal of this chapter is to describe main achievements in this ample area,
trying to follow the historical order. We consider mainly the case of a homogeneous (uniform) 
magnetic field. The papers related to inhomogeneous fields are cited in section \ref{sec-inhom}.

The main physical system under study is 
 a quantum spinless particle with mass $M$ and charge $e$, moving
in the $xy$-plane under the action of a uniform magnetic  field 
${\bf{H}}=\left(0,0,H_0\right) =\mbox{rot}{\bf A}({\bf r})$,  
directed along $z$-axis. Some oscillator-like potential 
$V(x,y) = k_1 x^2 + k_2 y^2$ can be also added. The Hamiltonian reads
\be
\hat{H} = \frac1{2M}\left[\hat{\bf p} -\frac{e}{c}{\bf A}(x,y)\right]^2 +V(x,y).
\label{Ham}
\ee

\subsection{The early history: before 1968}

Solutions of the stationary Schr\"odinger equation $\hat{H}\psi(x,y)=E\psi$
with Hamiltonian (\ref{Ham})
were found for the first time by Fock \cite{Fock28}  for the so called ``circular'' (or ``symmetric'') gauge
of the vector potential ${\bf A}_s =[{\bf H}\times{\bf r}]/2= (H_0/2)(-y,x)$
and $V(x,y)= M\omega_0^2\left(x^2 +y^2\right)/2$.
In this case Hamiltonian (\ref{Ham}) can be written also as
\be
\hat{H} = \frac1{2M}\hat{\bf p}^2 + \frac{M}{2}\tilde\omega^2 \hat{\bf r}^2 
- \omega_L \hat{L}_z,  \quad \omega_L =\frac{eH_0}{2Mc}, \quad \tilde\omega^2 = \omega_0^2 +\omega_L^2,
\label{Ham2}
\ee
where $\hat{L}_z=\hat{x}\hat{p}_y -\hat{y}\hat{p}_x$ is the canonical angular momentum operator.
Normalized orthogonal solutions in polar coordinates can be expressed in terms of the 
generalized Laguerre polynomials (hereafter $\mu \equiv M\tilde\omega/\hbar$):
\be
\psi_{n_r l}(r,\vf) = \sqrt{\frac{\mu n_r!}{\pi \left(n_r +|l|\right)!}}
\left(\mu r^2\right)^{|l|/2}L_{n_r}^{(|l|)}\left(\mu r^2\right) \exp\left(-\,\frac{\mu}{2} r^2 +i l\vf\right).
\label{psiFock}
\ee
The radial and angular momentum quantum numbers determine the energy levels
\be
E_{n_r l} =\hbar\tilde\omega\left(1 +|l| +2n_r\right) -\hbar\omega_L l, \qquad
n_r=0,1,2,\ldots, \quad l=0,\pm 1,\pm2, \ldots\,.
\label{Emag}
\ee
This problem was analyzed by Darwin in \cite{Darwin31}.
The special case of $V=0$ was solved also by Page \cite{Page30},
although the Laguerre polynomial structure of the solutions was not recognized by him.

Landau \cite{Land30} obtained solutions in terms of the Hermite polynomials for $V=0$,
choosing the gauge ${\bf A} = H_0(-y,0)$ (called now as ``Landau gauge'').
The remarkable feature of solutions with $V=0$ (the ``free particle'' case) 
is the {\em infinite degeneracy\/} of the energy spectrum,
which results in many interesting consequences.

The first example of non-spreading Gaussian packets in the presence of a homogeneous magnetic field
(with $V=0$) was given 
at the dawn of quantum mechanics by Darwin \cite{Darwin}.
The quantum mechanical propagator in this case was obtained by Kennard \cite{Kenn}. 
Another interesting example was given later by Husimi \cite{Hus1}.
The role of constants of motion was emphasized and elucidated by
Johnson and Lippmann \cite{JL49}.

\subsection{Main achievements since 1968: the content of this chapter}

The first coherent states of a charged particle in a uniform stationary magnetic field were 
constructed by Malkin and Man'ko \cite{MM69} as straightforward generalizations 
of the Glauber coherent states \cite{Glauber63} of a one-dimensional harmonic oscillator
to the case of two spatial dimensions. 
These states are discussed in Section \ref{sec-MM},
together with similar states introduced by other authors a little later. The further
generalizations: to the case of a time-dependent magnetic field and for 
relativistic particles, described by the Klein--Gordon and Dirac equations, -- are also
considered in this section.

From the modern point of view,
the Malkin--Man'ko coherent states (MMCS) can be considered as the simplest special case of a large family of
coherent states introduced by Klauder \cite{Klauder63}
(and later by Perelomov \cite{Perelomov}). These states
have the form $\exp\left[i\hat{Q}\right]|f\rangle$, where $|f\rangle$ is some ``fiducial'' state
 and $\hat{Q}$ is some linear combination of generators of a Lie group.
Namely, the MMCS are obtained from $|f\rangle =|0\rangle$ (the vacuum state) by applying to it the operator
$\exp\left[i\hat{Q}_1\right]$, where $\hat{Q}_1$ is a  linear combination of
 the annihilation and creation operators. 
The next step is to act on the coherent (or some other) states  by the operator
$\exp\left[i\hat{Q}_2\right]$, where $\hat{Q}_2$ is some  {\em quadratic\/}
form of the annihilation and creation operators. 
Such states became very popular under the name of ``squeezed'' states since 1980s \cite{Walls83}, 
although they were considered,
as a matter of fact, much earlier  \cite{Kenn,Hus1,Friedrichs53,Pleb55}.
These states have the form of more or less generic {\em Gaussian
wave packets} (in the case of fiducial coherent states). Squeezed states of non-relativistic
particles in a homogeneous magnetic field are considered in Section \ref{sec-sqz}.
A special attention there is paid to the so called ``geometrical'' squeezed states and the
Gaussian packets with a fixed mean value of the angular momentum.

Non-Gaussian wave packets are another wide family of quantum superpositions.
They can be created using different procedures. One of them is to apply the Klauder scheme
to non-vacuum (non-coherent) fiducial states. This line takes its origin from the displaced
Fock states of Plebansky \cite{Pleb54}. The second direction is to look for eigenstates
of squares or products of the annihilation operators. It takes its origin from the
paper by Barut and Girardello \cite{BG}. One of the simplest examples of such states
are {\em even and odd coherent states\/} \cite{evod}, 
which are eigenstates of the operators $\hat{a}^2$.
Specific features of analogs of these states for two space dimensions
in the presence of a magnetic field (including some inhomogeneous fields) are discussed
in Section \ref{sec-nonG}. Concrete subfamilies of non-Gaussian states, considered there,
include ``partially coherent'' and ``semi-coherent'' states, ``photon-added states'',
various kinds of ``nonlinear coherent states'', ``supersymmetric coherent states'',
and some others.

\section{Basic equations and their integrals of motion}

The equations of motion for a free charged particle in a homogeneous
magnetic field are as follows (they are the same both in the
classical case and for the Heisenberg operators in the quantum case): 
\be
\ddot{x} = \omega_c \dot{y}, \qquad
\ddot{y} = -\omega_c \dot{x},
\label{ddotxy}
\ee
where $\omega_c=eH_0/Mc$  is the cyclotron frequency.
If this frequency does not depend on time, then a consequence of
Eqs. (\ref{ddotxy}) is the existence of linear integrals of motion
\begin{equation} 
    {X}= {x}+ {\pi}_{y}/({M \omega_c}),  \qquad
    {Y}={y}- {\pi}_{x}/({M \omega_c}),
		\label{XY}
\end{equation}
where   
$\mbox{\boldmath$\pi$}={\bf{p}}-e{\bf{A}}/c =M{\bf v}$ is the {\it kinetic 
momentum}. These constants of motion are 
nothing but the coordinates of the center of a circle which the particle 
rotates around. Such an interpretation was crucial for the derivation of the
famous formula of the Landau diamagnetism \cite{Land30}. Later on, the
 significance of integrals of motion (\ref{XY}) was emphasized in Refs. \cite{JL49,32,33,34,Mielnik11}.
In particular, they are important for the construction of coherent and squeezed states.
The coordinates of the relative motion (with respect to the center of trajectory)
are proportional to the kinetic momenta. In the operator form they can be written as
\begin{equation} 
\widehat{\xi}=\widehat{x}-\widehat{X}=-\widehat{\pi}_{y}/({M \omega_c}), \qquad 
\widehat{\eta}=\widehat{y}-\widehat{Y}= \widehat{\pi}_{x}/({M \omega_c}). 
\label{5-6}
\end{equation}
The kinetic momenta operators do not commute:
$
\left[\hat{\pi}_x, \hat{\pi}_y\right] = i\hbar M\omega_c$. 
Consequently, the following commutation relations hold:
\begin{equation} 
\left[\widehat{\xi},\widehat{\eta}\right]= - \left[ 
\widehat{X}, \widehat{Y} \right]=\frac{i\hbar}{M\omega_c},\qquad
\left[\widehat{\xi},\widehat{X}\right]=\left[\widehat{\xi},\widehat{Y}\right]=
\left[\widehat{\eta},\widehat{X}\right]=\left[\widehat{\eta},\widehat{Y}
\right]=0.
\label{maincomm}
\end{equation}

Another consequence of Eqs. (\ref{ddotxy}) is the the existence of the quadratic
integral of motion, which can be considered as the generalized angular momentum:
\be
L = x\pi_y - y\pi_x + M\omega_c \left(x^2 + y^2\right)/2.
\label{L-pix}
\ee
It coincides formally with the canonical angular momentum
$
L_{can} =x p_y - y p_x $
in the case of ``circular'' gauge of the vector potential. 
The Hamiltonian (\ref{Ham}) and the angular momentum  (\ref{L-pix})
can be written in terms of ``geometric'' coordinates as follows:
\be
{H} =  M\omega_c^2\left(\xi^2 +\eta^2\right)/2,
\qquad
{L} =  M\omega_c\left({X}^2 +{Y}^2 -\xi^2 -\eta^2\right)/2.
\label{HL}
\ee
In addition, the Hamiltonian is proportional to the ``intrinsic'' angular momentum
\be
J = \xi \pi_y - \eta \pi_x = -2H/\omega_c,
\ee
which is important for constructing coherent states  \cite{Kowalski05}.

\subsection{Annihilation and creation operators in the magnetic field}

The main ingredients for building coherent states of the Glauber type \cite{Glauber63} 
are the annihilation and creation operators,
satisfying the boson commutation relations 
$\left[\hat{a},\hat{a}^{\dagger}\right] = \left[\hat{b},\hat{b}^{\dagger}\right] =1$.
There are many possibilities to construct such operators as linear combinations of four
operators $\hat{x}$, $\hat{y}$, $\hat{p}_x$, $\hat{p}_y$. But in view of commutation
relations (\ref{maincomm}), the most natural choice seems to be  \cite{MM69}
\begin{equation} 
\widehat{b}=\sqrt{\frac{M \omega_c}{2\hbar}}\left(
\widehat{X}-i\widehat{Y}\right),\qquad 
\widehat{a}=\sqrt{\frac{M \omega_c}
{2\hbar}}\left(\hat{\eta}-i\hat{\xi}\right)  = \frac{\hat\pi_x +i\hat\pi_y}{\sqrt{2\hbar M\omega_c}}.
\label{9-10}
\end{equation}
We assume hereafter that $\omega_c >0$.
Then the following relations hold:
\be
\hat{H} = \hbar\omega_c\left(\hat{a}^{\dagger}\hat{a} + \frac12 \right),
\qquad
\hat{L} = \hbar\left(\hat{b}^{\dagger}\hat{b} -\hat{a}^{\dagger}\hat{a}  \right),
\qquad
\left[\hat{L},\hat{a}\hat{b}\right] = \left[\hat{L},\hat{a}^{\dagger}\hat{b}^{\dagger}\right]=0.
\label{HLab}
\ee
It is worth noting that the sign of $\hat{L}$ should be inverted if $\omega_c <0$.
[One should be careful with the sign of cyclotron frequency $\omega_c$: many confusions
appear in various papers due to the negative sign of the electron charge.
In view of equations of motion (\ref{ddotxy}), if $\omega_c <0$, then one should replace $\omega_c$
with $|\omega_c|$ and make the rotation by $90$ degrees in the coordinate plane: $x \to y$, 
$y \to -x$. ]

\section{Malkin--Man'ko coherent states}
\label{sec-MM}

Malkin and Man'ko \cite{MM69} have introduced the two-dimensional coherent states 
$|\alpha,\beta\rangle$, which are common eigenstates of operators 
$\hat{a}$ and $\hat{b}$ defined by relations (\ref{9-10}):
\be
\hat{a}|\alpha,\beta\rangle = \alpha|\alpha,\beta\rangle, \qquad 
\hat{b}|\alpha,\beta\rangle = \beta|\alpha,\beta\rangle.
\label{abalb}
\ee
They found the following expression for the function 
$\Phi_{\alpha\beta}(x,y)= \langle x,y|\alpha,\beta\rangle$: 
\begin{equation}
        \Phi_{\alpha\beta}^{(MM)}=\sqrt{\frac{M \omega_c}{2 \pi \hbar}}
	\exp \left[- \zeta \zeta^{*}+ \sqrt{2} \beta \zeta+
	i \sqrt{2} \alpha \zeta^{*}-
	i \alpha \beta
        -\frac12\left(|\alpha|^{2}+|\beta|^{2}\right) \right],
\label{MMcoh}				
\end{equation}
\be
        \zeta=\sqrt{\frac{M \omega_c}{4\hbar}}(x+iy) , \qquad
\hat{a} = -\,\frac{i}{\sqrt2}\left(\zeta +\frac{\partial}{\partial \zeta^*}\right),
\quad		
\hat{b} = \frac1{\sqrt2}\left(\zeta^* +\frac{\partial}{\partial \zeta}\right).
			\label{def-zeta}
\ee

Perhaps, it is worth noting here, that if one has a vector operator
$\hat{\bf A} = \left(\hat{a}_1,...\hat{a}_N\right)$, whose components satisfy the relations 
$\left[\hat{a}_j, \hat{a}_k^{\dagger}\right] = \delta_{jk}$ and $ \left[\hat{a}_j, \hat{a}_k\right] =0$,
then the most simple way to obtain the eigenfunction of $\hat{\bf A}$
is to solve the following set of $2N$ coupled equations for the function
$f({\bf r};{\bf \alpha})=\langle {\bf r}|{\bf \alpha}\rangle
\exp\left(|{\bf \alpha}|^2/2\right)$
\cite{183vol,book3}:
\be
\hat{\bf A}f({\bf r};{\bf \alpha})={\bf \alpha} f({\bf r};{\bf \alpha}),
\quad
\hat{\bf A}^{\dagger}f({\bf r};{\bf \alpha})=
{\partial f({\bf r};{\bf \alpha})}/{\partial{\bf \alpha}}.
\quad
\label{eq-coh}
\ee

 The same coherent states (\ref{abalb})-(\ref{MMcoh}) were constructed a little later by Feldman and Kahn \cite{FeKa70}.
In both papers, \cite{MM69} and \cite{FeKa70}, the circular gauge of the vector potential was used.
An arbitrary choice of gauge was considered by Tam \cite{Tam71}. 
However, the transformation from the gauge ${\bf A}_s({\bf r})$ to another gauge
$
{\bf A}({\bf r}) = {\bf A}_s({\bf r}) + \nabla g({\bf r})$
is quite simple:
$
\Phi_{\alpha\beta}^{(g)}(x,y) = \Phi_{\alpha\beta}^{(MM)}(x,y) \exp\left[ie g({\bf r})/(\hbar c)\right]$.
The Landau gauge was used in \cite{DMM72,Boon75}.
The transition to the free particle (the zero magnetic field limit) in the coherent states (\ref{MMcoh})
was studied in Refs. \cite{Var84,Rowe91}. Some generalizations were considered recently in \cite{Lemo16}.

The decomposition of the coherent state (\ref{MMcoh}) over eigenstates of operators $\hat{a}^{\dagger}\hat{a}$ 
and $\hat{b}^{\dagger}\hat{b}$ has the standard form
\be
|\alpha,\beta\rangle =\exp\left[-\left(|\alpha|^2 +|\beta|^2\right)/2\right]
\sum_{n,m=0}^{\infty}\frac{\alpha^n \beta^m}{\sqrt{n! m!}}|n,m\rangle,
\label{alnm}
\ee
\be
\hat{a}^{\dagger}\hat{a}|n,m\rangle =n|n,m\rangle, \qquad  \hat{b}^{\dagger}\hat{b}|n,m\rangle = m|n,m\rangle,
\qquad \hat{L}|n,m\rangle = \hbar(m-n)|n,m\rangle.
\label{nmstates}
\ee
Mean values of the energy and angular momentum are given by the formulas
\be
\langle E \rangle = \hbar\omega_c\left(|\alpha|^2 +1/2\right), \qquad
\langle L \rangle = \hbar\left(|\beta|^2 -|\alpha|^2 \right).
\label{EL-MM}
\ee
Fluctuations of the angular momentum are decribed by the variance
\be
\sigma_{L} =\langle \hat{L}^2 \rangle - \langle \hat{L} \rangle^2 =
\hbar^2 \left(|\beta|^2 +|\alpha|^2 \right),
\ee
and they can be very large if $|\beta|=|\alpha| \gg 1$, although $\langle L \rangle=0$
for such states.
In view of Eq. (\ref{HL}), the quantum numbers $n$ and $m$ give quantized eigenvalues of
operators of squares of the relative radius $\xi^2 +\eta^2$ and the center of orbit radius
$X^2 +Y^2$ \cite{JL49}:
\be
\left(\xi^2 +\eta^2\right)_n = \frac{\hbar}{M\omega_c}(2n+1),\qquad
\left(X^2 +Y^2\right)_m = \frac{\hbar}{M\omega_c}(2m+1).
\label{eigRR}
\ee
The first equality means that the magnetic flux through the 
circular orbit of a charged particle is quantized: $\Phi_n \equiv H_0\pi\left(\xi^2 +\eta^2\right)_n
=(hc/e)(n+1/2)$.

\subsection{Applications and similar constructions}

Since coherent states form a complete set, they can be used to find the quantum propagator
$G({\bf r};{\bf r}';t) \equiv \langle {\bf r}|\hat{U}(t)|{\bf r}'\rangle$ by means of a simple
Gaussian integration. Using this approach, one does need any knowledge of energy eigenstates
and the bilinear summation formulas for the orthogonal polynomials (such as Mehler's formula).
If $\hat{U}(t)= \exp(-i\hat{H}t/\hbar)$, then the Wick rotation $t=-i\beta\hbar$ yields
immediately the equilibrium density matrix $\hat\rho_{eq}=\exp(-\beta\hat{H})$. 
In turn, the knowledge of the density matrix enables one to calculate the equilibrium
statistical sum and all equilibrium average values: the mean energy, magnetization, etc.
This approach was used for the first time by Feldman and Kahn \cite{FeKa70}, 
who applied coherent states (\ref{MMcoh}) for a simple derivation of
the famous Landau formula for the diamagnetic susceptibility of a free particle in a
homogeneous magnetic field. Analogous calculations were performed in \cite{Gran74}. 
The emission of electromagnetic radiation in the transition between two coherent states in the
homogeneous magnetic field (synchrotron radiation) was calculated in Ref. \cite{Liza75,Walls75,Bobrov84}.
In particular, it was concluded in \cite{Walls75} that
``the average energy absorbed from the radiation field when the cyclotron oscillators are
initially in an $n$ quantum state is considerably less than when the initial state is a coherent
superposition of number states.''
MMCS were used to calculate fluctuations of thermomagnetic currents in \cite{Agayeva85}.
The oscillation (de Haas--van Alphen) effects 
in the magnetization were considered in the framework of the approach based on  the
coherent states  by Pavlov {\em et al\/}
 \cite{Pavlov90, Pavlov91,Pavlov91a}. This subject was discussed in detail in \cite{183vol}
and later in Ref. \cite{Jellal01}.
For other applications see Refs. \cite{Cristofano95,Cristofano96,Schuch03},
where the states similar to the Malkin--Man'ko ones were used, in particular, 
in connection with the problem of dissipation in the presence of a homogeneous magnetic field.
Further generalizations (magneto-electric, bi-coherent and vector-coherent states) were considered in Refs. 
\cite{Hadji97,Ali05,Ali10}.
For the most recent publications see, e.g., \cite{Drigho17}.

\subsection{Time dependent coherent states}
\label{sec-td}

Coherent states in the case of time-dependent homogeneous magnetic fields were constructed
in \cite{DMM72,LR69,MMT69,MMT69a,MMT70,MM71,Holz70}.
The main idea belongs to Lewis and Riesenfeld \cite{LR69}, who showed that solutions
to the nonstationary problem can be found as eigenstates of some {\em time-dependent
integrals of motion} (quantum invariants), i.e., operators $\widehat {I}(t)$ satisfying the equation
$i\hbar\partial \widehat {I}/\partial t -[\widehat H\,,\,\widehat I ]=0$.
They found {\em quadratic\/} invariants (with respect to the coordinates and momenta
operators) for the quantum oscillator with a time-dependent frequency and the charged
particle in a time-dependent homogeneous magnetic field.
The next important step was made by Malkin, Man'ko and Trifonov \cite{MMT69,MMT69a,MMT70,MM71},
who showed that the calculations can be greatly simplified, if one looks for {\em linear\/}
integrals of motion. This idea was further developed in \cite{DMM72,Holz70,MMT73,DMM75}.

Hamiltonians (\ref{Ham}) and (\ref{Ham2}) are special cases of
the general quadratic Hamiltonian (without linear terms for the sake of simplicity)
\[
H=\frac12\sum_{j,k=1}^{2N}B_{jk}(t)q_jq_k \equiv \frac12{\bf q}B(t){\bf q},
\qquad B=\left\Vert\begin{array}{cc}
b_1&b_2\\
b_3&b_4\end{array}
\right\Vert,
\quad b_3 =\tilde{b}_2,
\]
where ${\bf q} =({\bf p},{\bf r})$ is the $2N$-dimensional vector, combining $N$-dimensional vectors ${\bf p}$
and ${\bf r}$, whereas $B$ is a $2N\times2N$ symmetric matrix consisting of $N\times N$ blocks
(the tilde means the transposed matrix).
Looking for $N$-dimensional linear integrals of motion in the form
$\hat{\bf A}(t)=\lambda_p(t)\hat{\bf p} +\lambda_r(t)\hat{\bf r}$, one can arrive  at the set
of coupled ordinary linear differential equations for the complex $N\times N$ matrices 
$\lambda_p$ and $\lambda_r$,
\be
\dot{\lambda}_p=\lambda_p b_3-\lambda_r b_1, \quad
\dot{\lambda}_r=\lambda_p b_4-\lambda_r b_2.
\label{eqlam}
\ee
To construct the time dependent coherent states, the initial conditions
should be chosen in such a way that $\hat{\bf A}(0) =\hat{\bf a}$, where $\hat{\bf a}$ is the vector
operator describing the selected set of initial annihilation operators.
Then the time-dependent coherent state, satisfying the equation 
$\hat{\bf A}|{\bf \alpha}\rangle = {\bf \alpha}|{\bf \alpha}\rangle$,
 has the following form in the coordinate representation
\cite{183vol,book3,DMM75} (here ${\bf \alpha}$ is the $N$-dimensional vector):
\be
\langle {\bf r}|{\bf \alpha}\rangle= \frac{\left(2\pi\hbar^2\right)^{-N/4}}
{\left(\det\lambda_p\right)^{1/2}}
\exp\left(-\frac i{2\hbar} {\bf r}\lambda_p^{-1}\lambda_r{\bf r}
+\frac{i}{\hbar} {\bf r}\lambda_p^{-1}{\bf \alpha}
+\frac12\,
{\bf \alpha}\lambda_p^*\lambda_p^{-1}{\bf \alpha}
-\frac12 
|{\bf \alpha}|^2
\right) .
\label{coh-N}
\ee

For Hamiltonian (\ref{Ham}) or (\ref{Ham2}), we have $b_1 =M^{-1} E_2$, where $E_2$ is the $2\times2$
unit matrix. Moreover, $b_4=M \tilde{b}_2 b_2$ in the absence of an additional potential, while
 the structure of matrix $b_2$   depends on the choice of gauge of the vector potential.  
For the symmetric (S) and Landau (L) gauges we have,  respectively,
\[
 b_2^{(S)} = \Omega_S(t)\left\Vert
\begin{array}{cc}
0 & 1
\\
-1 & 0
\end{array}
\right\Vert, 
\quad
b_2^{(L)} = \Omega_L(t)\left\Vert
\begin{array}{cc}
0 & 1
\\
0 & 0
\end{array}
\right\Vert, 
\qquad
\Omega_L(t) = 2\Omega_S(t) =\omega_c(t).
\]
It is easy to verify that in the case of symmetric gauge, due to the property $b_3= -b_2$,
the solutions to equations (\ref{eqlam}) can be found in the form
\[
\lambda_p = \vep(t) F U(t), \quad \lambda_r = -M\dot\vep(t) F U(t), \quad
U(t) = \exp\left[\frac12 \int_0^t \left(b_3 -b_2\right)d\tau\right],
\]
where $F$ can be an arbitrary constant matrix and the scalar function $\vep(t)$ can be any
solution to the classical equation of motion
for the harmonic oscillator with the time-dependente frequency $\Omega=\Omega_{S}(t)$: 
\be
 \ddot{\varepsilon}+\Omega^2(t) \varepsilon=0.
\label{eqvep}
\ee
The explicit form of the unitary matrix $U(t)$ is as follows,
\[
U(t) = \left\Vert
\begin{array}{cc}
\cos(\phi) & -\sin(\phi)
\\
\sin(\phi) & \cos(\phi)
\end{array}
\right\Vert,
\qquad
\phi = \int_0^t \Omega_S(\tau)d\tau.
\]
To construct the coherent states, it is convenient to choose the complex solution to Eq. (\ref{eqvep}), 
satisfying the condition
\be
 \dot{\varepsilon}\varepsilon^{*}-\dot{\varepsilon}^{*} \varepsilon=2i.
\label{Wr}
\ee
If $\Omega=const >0$, then the required solution has the form $\vep(t)= \Omega^{-1/2}\exp(i\Omega t)$.
In this case,  $\hat{\bf A}(0) =\left(\hat{a}, \hat{b}\right)$ [where $\hat{a}$ and
$\hat{b}$ are given by Eq. (\ref{9-10})] if 
\[
F= (2\sqrt{M\hbar})^{-1} \left\Vert
\begin{array}{cc}
1 & i
\\
i & 1
\end{array}
\right\Vert, 
\qquad
FU = (2\sqrt{M\hbar})^{-1} \left\Vert
\begin{array}{cc}
e^{i\phi} & i e^{i\phi}
\\
i e^{-i\phi} &  e^{-i\phi}
\end{array}
\right\Vert.
\]
Then general formula (\ref{coh-N}) yields the following generalization of (\ref{MMcoh}) to the case of
the time-dependent symmetric gauge of the vector potential 
[here $\tilde\zeta \equiv \sqrt{{M}/{\hbar}}\,(x+iy)$]:
\be
\langle x,y|\alpha,\beta\rangle = \left(\pi\hbar\vep^2/M\right)^{-1/2} \exp \left[
\frac{i\dot\vep}{2\vep}|\tilde\zeta|^2 +\vep^{-1}\left(i\alpha\,e^{-i\phi}\,\tilde\zeta^* 
+ \beta \,e^{i\phi}\,\tilde\zeta \right) 
-i\alpha\beta \vep^*/\vep -\frac12\left(|\alpha|^2 +|\beta|^2\right)\right]. 
\label{coh-sym}
\ee
Formula (\ref{coh-sym}) was obtained (in slightly different forms) in Refs. \cite{MMT69,MMT69a,MMT70,MM71}.
Similar results were found later, e.g., in \cite{Abdalla88}.
Additional time-dependent homogeneous electric fields were considered in \cite{DMM72,MM71,MM71a}.
Explicit expressions for the function $\vep(t)$ in some special cases were found in \cite{Agayeva80}
(see also \cite{book3} for the list of known explicit solutions).
Integrals of motion and their eigenfunctions in the case of non-commuting coordinate operators,
$\left[\hat{x},\hat{y}\right] = i\vartheta$, were studied in \cite{Fiore11} (for the symmetric gauge 
of the time-dependent vector potential).
The case of time-dependent Landau gauge is more complicated \cite{DMM72}.

Approximate quasiclassical packets, whose centers move along classical trajectories in arbitrary (inhomogeneous)
electro-magnetic fields, were studied in Refs. \cite{Bag-traj82,DodOs90}. The case of homogeneous 
magnetic field was considered in the frame of this approach in Ref. \cite{Belov90}.
More general constructions were considered in \cite{Mantoiu11}.
A method of generation of electron Gaussian coherent packets was proposed in \cite{Ryu16}.

\subsection{Relativistic coherent states on the null plane}

The main difficulty for constructing coherent states in the relativistic case 
(for the Klein--Gordon or Dirac equations)
is the non-equidistant energy spectrum. 
For example, the spectrum of the Dirac particle in the homogeneous magnetic field, first obtained
by Rabi \cite{Rabi28}, has the form
$E_n = \pm \sqrt{M^2c^4 + p_z^2 c^2 + 2Mc^2 \hbar \omega_c(n+1)}$.
For this reason, superpositions defined as in Eq. (\ref{alnm}) do not preserve their
form with time. A possibility to overcome this difficulty was found in
\cite{Bagrov75,DMM76,Bagrov76}. Let us consider, following \cite{DMM76}, the Klein--Gordon
equation for a charged particle of mass $M$ in uniform magnetic field $B$, directed along the $z$ axis. 
Introducing the ``null plane operators''
$\hat\xi_3 = \hat{p}_0 -\hat{p}_z$ and  $\hat\eta_4 = \left(\hat{p}_0 +\hat{p}_z\right)/2$
(with $\hat{p}_0=i\partial/\partial t$), one can rewrite
the equation in the form
\be
\left[ - \hat\xi_3 \hat\eta_4 +\frac12\left(\hat{\pi}_x^2 +\hat{\pi}_y^2\right) 
+ \frac12 M^2 \right] \psi =0.
\label{KG}
\ee
Here we assume $e=c=\hbar=1$ and use the pseudo-euclidian metric with $g^{00} = -g^{aa}=1$, where $a=1,2,3$.

The operator $\hat\xi_3 \equiv \hat{I}$ is the integral of the motion for Eq. (\ref{KG}). 
Therefore in the space of eigenfunctions of this operator with the fixed eigenvalue I, this equation 
can be considered as the usual Schr\"odinger equation, if one introduces the ``new time'' 
$s = (t - z)/I$. Then one can write $\hat\xi_3 \hat\eta_4 = i\partial/\partial s$,
so that the integrals of the motion, generating coherent states, can be chosen as 
$\hat{A} =\hat{a}\exp(i\omega_c s)$ and $\hat{b}$, where $\hat{a}$ and $\hat{b}$ are given by Eq. (\ref{9-10}).
Their eigenstates are the Gaussian packets with respect to the transverse coordinates
(in the circular gauge of the vector potential) \cite{DMM76}:
\beqn
\psi_{\alpha\beta I}(x,y,z,t) &=& (2\pi)^{-2} \sqrt{B} \exp\left\{-\,\frac{is}{2}\left(B+M^2\right)
-\frac12 I(z+t) -\frac{B}{4}\left(x^2 +y^2\right) \right. \nonumber \\
&+& \left.  \sqrt{\frac{B}{2}} \left[\alpha_s (x-iy) +\beta(x+iy)\right] -\alpha_s \beta 
-\frac12\left(|\alpha|^2 +|\beta|^2\right) \right\},
\label{cohrel}
\eeqn
where $\alpha_s =\alpha \exp(-iBs)$.
The case of homogeneous electric field and the field of a plane wave was studied in \cite{Bagrov76} 
using the ``null plane'' formalism. Further developments in this direction can be found in 
\cite{Ternov83,Colavita14}.

Approximate coherent states of the Dirac particle in a uniform magnetic field were constructed in \cite{Filippov98}
in the case of high mean excitation quantum numbers, 
$\overline{n}=\langle \hat{a}^{\dagger}\hat{a}\rangle \gg 1$, 
when the energy spectrum can be considered as effectively equidistant.
Quasiclassical ``trajectory coherent'' states for a charged relativistic particle obeying the Klein--Gordon equation 
 were considered 
in  \cite{Bagrov82,Belov88,Gritsenko99}, whereas the case of the Dirac particle was studied in Ref. \cite{Bagrov93}.
Gaussian wave packets for the Klein--Gordon particle in the Foldy representation 
were constructed in \cite{Mostaf06}.
Coherent-like superpositions of energy states for the Dirac particle in a uniform magnetic field
were considered in \cite{Bermudez07}, using some analogies with the famous Jaynes--Cummings model of quantum 
optics. The dynamics of such packets was studied also in \cite{Demik12}.

\section{Squeezed states and Gaussian packets}
\label{sec-sqz}

Time dependent Gaussian packets of Refs. \cite{DMM72,MMT69,MMT69a,MMT70,MM71,Holz70}, discussed in Sec. \ref{sec-td},
can be interpreted nowadays as two-dimensional squeezed states.
Such packets were studied also in\cite{KimWei73}. 
However, one of the first examples was given in 1953 by Husimi \cite{Hus1}, who found the following time-dependent
packets for the circular gauge of the constant magnetic field (in dimensionless units):
\be
\psi({\bf r},t; {\bf a}, \beta) = \frac{\sqrt{\sinh(2\beta)/(2\pi)}}{\sinh(\beta+it)}
\exp\left\{ -\frac12\coth(\beta+it)({\bf r} - {\bf a})^2 
 -[{\bf r}\times{\bf a}]_z 
-a^2 \coth(2\beta) \right\}.
\label{Husmag}
\ee

Explicit words ``squeezed states in the magnetic field'' were used, e.g., in papers 
\cite{Bechler88,Jan89,Abdmag91,Kov,Baseia92,BasMiMo,DelMi}.
Evolution of squeezed states in the presence of a magnetic field was considered in \cite{Santos09}.
The most general construction $\exp\left[i\hat{Q}_2\right]|{\bf \alpha}\rangle$ for the particle in a magnetic field was
studied in detail in \cite{DKurM176} under the name ``correlated coherent states''.
Similar states, defined as common eigenstates of the operators 
$\hat{A} =\left(\hat{a}-\lambda \hat{b}^{\dagger}\right)/\sqrt{1-|\lambda^2}$
and $\hat{B} =\left(\hat{b}-\lambda \hat{a}^{\dagger}\right)/\sqrt{1-|\lambda^2}$,
were studied recently in \cite{Dehghani12}.

\subsection{``Geometrical'' squeezed states}

In many papers \cite{Bechler88,Jan89,Abdmag91,Kov,Baseia92,BasMiMo,DelMi,Santos09},
 the squeezing phenomena were considered with respect
to the {\em canonical pairs\/} of variables, such as $x,p_x$ and $y,p_y$.
However, the physical meaning of the numerous formulas for the variances of these variables
is not quite clear. Therefore it was suggested in \cite{DKurM176,Arag,GSQ} to analyze the variances in the
pairs $(X,Y)$ (the center of orbit coordinates) and $(\xi,\eta)$ (the relative motion coordinates).
The states possessing variances of 
any element of the pairs $(X,Y)$ or $(\xi,\eta)$ less than $\hbar/2m\omega_0$
were named ``geometrical squeezed states'' (GSS) in \cite{GSQ}, in order to emphasize that all
the observables $(X,Y,\xi,\eta)$ have the meaning of 
coordinates in the usual (``geometrical'') space, and not in the phase space.
The squeezed states with respect to the $X-Y$ pair were constructed in \cite{OzaShe} as
common eigenstates of the Hamiltonian and the operator $\hat{X}\cos(\Phi) + \hat{Y}\sin(\Phi)$,
where $\Phi$ is a complex parameter with negative imaginary part.
Applications of the squeezed states in the magnetic field to charged electron--hole systems
were considered in \cite{Dzyub01,Dzyub01a}.

An interesting problem raised in \cite{GSQ} is how one could create GSS, starting from coherent states of the 
Malkin--Man'ko type?
For the single-mode systems such a problem can be solved effectively by using quadratic Hamiltonians with time dependent 
coefficients \cite{183vol,book3,Takahasi65,DM94}.
But whether this can be done using time-dependent magnetic fields in two dimensions?
It appears that the answer depends on the choice of time-dependent gauge (or, from a more physical
point of view, on the structure of the induced electric field).

Let us suppose that the magnetic field varies in time at $0<t<\tau$, 
assuming the same constant value $H_0$ for $t<0$ and for $t>\tau$. In such a 
case, it is possible to achieve any degree of squeezing in $x-p_x$ or $y-p_y$ pairs, starting from 
any coherent state \cite{Bechler88,Jan89,Abdmag91,Kov,Baseia92,BasMiMo}. However, the situation 
becomes quite different when one considers squeezing in the $X-Y$ or $\xi-\eta$ pairs.

The strength of quantum fluctuations is characterized usually by the values of the variances
    $ \sigma_{\alpha\beta}=\frac{1}{2}\langle\widehat{\alpha}
\widehat{\beta}+ \widehat{\beta}\widehat{\alpha}\rangle-
\langle\widehat{\alpha}\rangle\langle\widehat{\beta}\rangle$ 
(where $\alpha,\beta=X,Y,\xi,\eta$), combined into the variance matrix
$\mbox{\boldmath$\sigma$}=\Vert\sigma_{\alpha\beta}\Vert$. If the magnetic 
field does not depend on time, the Hamiltonian 
$\widehat{H}=\widehat{\mbox{\boldmath$\pi$}}^2/2m=
m\omega_c^{2}\left(\widehat{\xi}^{2}+\widehat{\eta}^2\right)/2$
does not contain operators $\widehat{X}$ and $\widehat{Y}$.
Then the $X-Y$ variances are constant in time, while the $\xi-\eta$ variances 
perform harmonic oscillations. For example, for $t>\tau$,
\[\sigma_{\xi\xi}(t)=\sigma_{\xi\xi}(\tau )\cos^2\left(\omega [t-
\tau ]\right)+\sigma_{\eta\eta}(\tau )\sin^2\left(\omega [t-\tau 
]\right)+\sigma_{\xi\eta}(\tau )\sin\left(2\omega [t-\tau ]\right).\]
It is not difficult to find the minimum of this expression as function of $t$:
\begin{equation}
\sigma_{\xi\xi}^{(\mbox{min})}=\frac 12\left[T-\sqrt {T^2-4d} \right],
\qquad 
T=\sigma_{\xi\xi}+\sigma_{\eta\eta}, \quad d=\sigma_{
\xi\xi}\sigma_{\eta\eta}-\sigma_{\xi\eta}^2.
\label{min}
\end{equation}
Formula (\ref{min}) was derived for the first time in Ref. \cite{Luks88}
under the name ``principal squeezing''.
The physical meaning of invariants $T$ and $d$ was clarified in Refs. \cite{183vol,book3,DR1}. 
$T$ is nothing but the double energy
of quantum fluctuations. It is conserved for time--independent Hamiltonians,
but it varies in time for the nonstationary Hamiltonians. As to the
parameter $d$, it is conserved in time for {\it any} nonstationary
(one--mode) Hamiltonian: the only restriction is that the Hamiltonian
must be {\it quadratic} with respect to operators $\widehat{\xi}$ and
$\widehat{\eta}$ \cite{183vol,book3,Dod-univ}. 
The importance of this parameter is explained by two
reasons. First, it satisfies the generalized uncertainty relation
$d\ge d_{\mbox{min}}\equiv(\hbar/2m\omega_c)^2$. Secondly, for the
{\it Gaussian} states described by the density matrix $\widehat{\varrho}$,
parameter $d$ characterizes {\it the degree of mixing} of the quantum
state, due to the relation \cite{167vol}
$\mbox{Tr}(\widehat{\varrho}^2)=\left(d_{\mbox{min}}/d\right)^{1/2}$ 
(we assume the normalization of the density matrix 
$\mbox{Tr}\widehat{\varrho}=1$). If we deal with a {\it one--mode} system, 
 $d(t)=d_{\mbox{min}}=const$ for any initial coherent state. Then an 
arbitrary parametric excitation of the mode yields 
$T(t)>T_{\mbox{in}}=2\sqrt{d_{\mbox{min}}}$, and the system occurs 
automatically in a squeezed state.
But in the case of interacting 
{\it multimode} systems the ``degree of mixing'' of every subsystem can 
{\it increase} upon the interaction, so that no squeezing will arise. 

To calculate the elements of the variance matrix at $t>0$, we introduce the 
operator vector $\widehat{{\bf{q}}}=\left(\widehat{X}, \widehat{Y}, 
\widehat{\xi}, \widehat{\eta}\right)$.
Since the Hamiltonian is {\it quadratic} with respect to the components
of vector $\widehat{{\bf{q}}}$, the Ehrenfest equations of motion for the
mean values of these components are {\em linear}. Consequently,
we have a linear relation 
$\langle\widehat{{\bf{q}}}\rangle(t)=\Lambda(t) \langle\widehat{{\bf{q}}}\rangle(0)$,
where $\Lambda(t)$ is some $4\times4$ symplectic matrix. Moreover,
 the initial variance matrix $\mbox{\boldmath$\sigma$}(0)$ 
and the final one $\mbox{\boldmath$\sigma$}(t)$ are related by means of the same matrix $\Lambda(t)$ as follows:
$ \mbox{\boldmath$\sigma$}(t)=\Lambda(t)\mbox
{\boldmath$\sigma$}(0)\Lambda^{T}(t)$. Here
$\Lambda^{T}$ is the transposed matrix. 

Explicit expressions for the elements of matrix $\Lambda(t)$ in the special cases
of symmetric (``circular'') and Landau gauges were given in \cite{GSQ}.
The following formulas were obtained in the symmetric case for the initial coherent states:
\be
\sigma_{\xi\eta}=\sigma_{XY}=0, \quad
\sigma_{\xi\xi}=\sigma_{\eta\eta}=\sigma_{XX}=\sigma_{YY}
=\frac {\hbar}{8\omega_c^2m}\left[\omega_c^2|\varepsilon |^2+4|\dot\varepsilon |^2\right],
\label{sigsym}
\ee
where function $\varepsilon(t)$ is defined according to Eqs. (\ref{eqvep}) and (\ref{Wr}).
Therefore
\[
\sigma_{\xi\xi} \ge\frac {\hbar |\varepsilon\dot{\varepsilon }|}{2\omega_c m}\ge\frac {
\hbar\mbox{Im}(\varepsilon^{*}\dot{\varepsilon })}{2\omega_c m}=\frac {
\hbar}{2\omega_c m}.
\]
This result means that  a time--dependent magnetic field {\em with the 
axial symmetry of the accompanying vortex electric field\/} is not able to 
``squeeze'' an initial coherent state with respect to the $\xi -\eta$ and $X-Y$ pairs. 
This can be explained by an effective
``thermalization'' of the $\xi -\eta$ and $X-Y$ subsystems, since 
 we have $T=2\sqrt {d}$ and $d\ge d_{\mbox{min}}$
in the final state for each subsystem. 
Moreover, it can be shown that for any initial {\em squeezed\/} state the 
final {\em minimal variances\/} (in the sense of Eq. (\ref{min})) of both the 
guiding center and relative coordinates will be {\em greater\/} than the 
initial ones.

The case of time--dependent Landau gauge is more complicated \cite{DMM72}.
In this case one needs the solutions to the equation 
$ \ddot{\varepsilon}+\omega_c^2(t) \varepsilon=0$ (note the change in the
effective frequency, compared with the case of symmetric gauge),
satisfying the normalization condition (\ref{Wr})
(so that $\vep(t)=\omega_c^{-1/2} \exp(i\omega_c t)$ for $t<0$, when $\omega_c=const$). 
However, differently from Eq. (\ref{sigsym}), the (co)variances are not
determined completely by the instant values of functions $\vep(t)$ and $\dot\vep(t)$ only. 
The following additional functions of time appear in the final formulas:
\[
\sigma =\int_0^t\omega (\tau )\varepsilon (\tau )\,\mbox{d}\tau -{i}\omega_c^{-1/2},
\quad
s=\mbox{Im}\left(\varepsilon\sigma^{*} \right),
\quad
\kappa =\int_0^t\left[1-\omega (\tau )s(\tau )\right]\,\mbox{d}\tau .
\]
Introducing the  dimensionless variances 
$\widetilde{\sigma}_{\alpha\beta}\equiv 2m\omega_c\sigma_{\alpha\beta}/\hbar$, 
one can obtain the following formulas for the variances at $t>0$ 
(for the initial coherent state) \cite{GSQ}:
\[
\widetilde{\sigma}_{XX}=1+\left(\dot s-\omega_c\kappa\right)^2+\left
|\omega_c\sigma +\dot\varepsilon\right|^2/\omega_c,\quad
\widetilde{\sigma}_{YY}=1,\quad
\widetilde{\sigma}_{XY}=\dot {s}-\omega_c\kappa .
\]
\[
\widetilde{\sigma}_{\xi\xi}=\dot {s}^2+|\dot{\varepsilon }|^2/\omega_c, \quad
\widetilde{\sigma}_{\eta\eta}=\left(\omega_cs-1\right)^2+\omega_c|
\varepsilon |^2, \quad
\widetilde{\sigma}_{\xi\eta}=-\dot {s}\left(\omega_cs-1
\right)-\mbox{Re}\left(\dot\varepsilon\varepsilon^{*}\right).
\]

Two exactly solvable examples of function $\omega(t)$ were considered in \cite{GSQ}.
The first one was the {\em step--like variation}, when 
$\omega =\omega_c$ for $t<0$ and $t>\tau$, and $\omega =\omega_c\Theta 
=const$ for $0<t<\tau$. It was shown that squeezing in $\xi$-component is possible
for $\Theta <1$. But the minimal possible variance $\widetilde{\sigma}_{\xi\xi}$
was only $1/2$.
Another simple example was the $\delta$-kick of frequency: 
$\omega^2(t)=\omega_c^2+2\gamma\delta (t)$ with $\gamma >0$.
But in this case, again, the inequality 
$1> \widetilde{\sigma}_{\xi\xi}^{(\mbox{min})} > 1/2$ was shown to hold.
Therefore it was questioned in \cite{GSQ}, whether it is possible 
to achieve an {\it arbitrary strong} squeezing of the geometric coordinates?

Here we can show that the answer is positive for the $\xi-\eta$ pair in the case 
of periodical variation of the magnetic field in the form
$\omega(t) = \omega_c\left[1+2\gamma \cos(2\omega_c t)\right]$
(the standard example of the parametric resonance).
For $|\gamma| \ll 1$ we have an approximate solution 
(using, e.g., the method of averaging
over fast oscillations \cite{LL-mech} or the method of slowly varying
amplitudes and neglecting terms of the order of ${\cal O}(\gamma)$ in the amplitude coefficients)
\cite{DKN93}
\[
\vep(t) = \omega_c^{-1/2}\left[\cosh\left(\omega_c\gamma t\right) e^{i\omega_c t}
-i \sinh\left(\omega_c\gamma t\right) e^{-i\omega_c t}\right].
\]
Then
$\sigma(t) = \omega_c^{-1/2}\left[-i\cosh\left(\omega_c\gamma t\right) e^{i\omega_c t}
+ \sinh\left(\omega_c\gamma t\right) e^{-i\omega_c t}\right] = - \dot\vep(t)/\omega_c$,
so that $s(t) = 1/\omega_c$, $\kappa(t)=0$.
Consequently
$
\tilde\sigma_{\xi\xi}(t) = \cosh(2\omega_c\gamma t) +\sinh(2\omega_c\gamma t) \sin(2\omega_c t)$.
Hence any desired degree of squeezing can be obtained for the $\xi-\eta$ pair:
\[
\tilde\sigma_{\xi\xi}^{(min)}(t) = \exp\left(-2\omega_c \gamma t\right).
\]
However, no squeezing is observed for the $X-Y$ pair in this case:
$\widetilde{\sigma}_{XX}=\widetilde{\sigma}_{YY}=1$ and $\widetilde{\sigma}_{XY}=0$.
It could be interesting to find limitations on the minimal degrees of squeezing 
for the $\xi-\eta$ pair in the case of the ``intermediate'' gauge of the time dependent
vector potential ${\bf A}(t) = (H_0(t)/2)\left(-y(1+\beta),x(1-\beta)\right)$, varying the 
asymmetry parameter $\beta$ from $0$ to $1$.

\subsection{Minimum energy Gaussian packets with a fixed mean angular momentum
in a constant magnetic field}

Coherent states (\ref{MMcoh}) possess nonzero mean values (\ref{EL-MM}) of the angular momentum operator.
This means that charged quantum particles described by such wave functions perform some rotation in the 
$xy$ plane.
Rotated Gaussian packets or Gaussian packets in rotating frames were considered with different purposes in many papers 
\cite{GulVol80,GV80,Hac96,Wun02,Bracher11,Ross11,Ross14,Karlov15}.
The Gaussian packets possessing the minimal possible energy for a fixed 
mean values of the angular momentum operator were found in \cite{Drot}.
They have the following form in the polar coordinates $r,\vf$
(in this subsection we consider the case of a time-independent magnetic field, using the symmetrical 
gauge of the vector potential):
\beqn
\psi_{min}(r,\vf) &=& \sqrt{\mu/\pi}(1-\rho^2)^{1/4} 
\exp\left( -\,\frac{\mu}{2} r^2\left[ 1 + \rho \exp(2i\lambda\vf -i\lambda u) \right]
\right. \nonumber \\   &+& \left. 
\sqrt{\mu \left|{\cal L}_c\right|}\,r \left(\exp\left[i\lambda_c(\vf - v)\right] 
+\rho\exp[i\lambda(\vf +v -u)]\right) -{\Phi}
\right).
\label{G-polmin}
\eeqn
Here
\be
\rho =\sqrt{\frac{|{\cal L}_i|}{1+ |{\cal L}_i|}}, \qquad 
\Phi = \left|{\cal L}_c\right|\left[1 +\rho\cos(u-2v)\right]/2.
\ee
The equivalent form in the Cartesian coordinates $x=r\cos(\vf)$ and $y=r\sin(\vf)$ is
\be
\psi(x,y) = \sqrt{\mu/\pi}(1-\rho^2)^{1/4} \exp\left[-\mu\left( a x^2 +b xy + c y^2\right) +Fx +Gy -{\Phi} \right],
\label{psi}
\ee
\be
a=\frac12\left[1 +\rho\exp(-i\lambda u)\right], \quad
c=\frac12\left[1 -\rho\exp(-i\lambda u)\right], \quad
b= i\lambda\rho\exp(-i\lambda u),
\label{compl}
\ee
\be
F= \sqrt{\mu \left|{\cal L}_c\right|}\, \left\{\exp\left(-i\lambda_c v\right) + \rho\exp\left[i\lambda (v-u)\right]\right\},
\ee
\be
G= i\sqrt{\mu \left|{\cal L}_c\right|}\,\left\{\lambda_c\exp\left(-i\lambda_c v\right) 
+ \lambda\rho\exp\left[i\lambda (v-u)\right]\right\}.
\label{G}
\ee
To understand the meaning of parameters in Eqs. (\ref{G-polmin})-(\ref{G}), one should remember that 
 the first order mean values $\langle {\bf r}\rangle$ and $\langle {\bf p}\rangle$ are totally
independent from their (co)variances in the Gaussian states.
As a consequence, both the mean angular momentum and mean energy can be written as sums of two independent terms,
\[
\langle \hat{L}_z\rangle \equiv \hbar{\cal L} = \hbar\left({\cal L}_c + {\cal L}_i\right), 
\quad
{\cal E}= {\cal E}_c + {\cal E}_i,
\]
where ${\cal L}_c$ and ${\cal E}_c$ are determined completely by the mean values 
$\langle {\bf r}\rangle$ and $\langle {\bf p}\rangle$, 
whereas ${\cal L}_i$ and ${\cal E}_i$ depend only on fluctuations of these variables through their covariances.
For the fixed value ${\cal L}_c$, the ``classical'' part of energy attains the minimal value 
${\cal E}_c^{min}= \hbar\left|{\cal L}_c\right|$  for the points belonging to the circle 
$|\langle {\bf r}\rangle| =\sqrt{\mu \left|{\cal L}_c\right|}$.
Parameter $\lambda_c = \pm1$ determines the sign of ${\cal L}_c =\lambda_c\left|{\cal L}_c\right|$.
Similarly, ${\cal L}_i =\lambda\left|{\cal L}_i\right|$.
Parameters $u$ and $v$ are nothing but angles defining the orientation of the ellipse of constant probability
$|\psi(x,y)|^2=const$ and the position of the center of this ellipse in the circle 
$|\langle {\bf r}\rangle| =\sqrt{\mu \left|{\cal L}_c\right|}$.
The minor axis of this ellipse is inclined by angle $u/2$ with respect to $x$-axis. The major
and minor axes of the ellipse are proportional to $\left(1 \mp \rho\right)^{-1/2}$, 
and the ellipse eccentricity equals $\vep= \left[2\rho/\left(1+\rho\right)\right]^{1/2}$.
For the free particle in the uniform magnetic field with $\omega_L >0$ (without an additional 
oscillator potential), the angles vary with time as follows,
\be
u(t)=u_0 +2\omega_L t(\lambda-1), \quad v(t)=v_0 +  \omega_L t\left(\lambda_c-1\right).
\label{uvt-mag}
\ee

The mean energy of the packet (\ref{G-polmin}) equals
\be
{\cal E} = \hbar\omega_L\left[1 +\left|{\cal L}_i \right|(1-\lambda) + \left|{\cal L}_c \right|\left(1-\lambda_c\right) \right].
\ee
The absolute minimum
${\cal E}_{min} = \hbar\omega_L $ is achieved for all packets 
with $\lambda=\lambda_c =1$.
Such packets do not rotate at all, although they can possess arbitrary values of
``external'' (${\cal L}_c$) and ``internal'' (${\cal L}_i$) angular momenta.
Of course this is explained by the well known infinite degeneracy of energy eigenstates in the homogeneous
magnetic field.

The following expressions were found \cite{Drot} for the energy and angular momentum variances 
$\sigma_E =\langle \hat{H}^2\rangle - \langle \hat{H}\rangle^2$
and $\widetilde\sigma_L = \left(\langle \hat{L}^2\rangle - \langle \hat{L}\rangle^2\right)/\hbar^2$: 
\[
\widetilde\sigma_L =  \left|{\cal L}_c\right| + 2\left|{\cal L}_i\right| \left(1+\left|{\cal L}_i\right|\right)
+\left(1+\lambda\lambda_c\right) \left|{\cal L}_c\right| \left[ \left|{\cal L}_i\right| 
 -  \sqrt{\left|{\cal L}_i\right| \left(1+\left|{\cal L}_i\right|\right)}\cos(2w) \right],
\]
\beqnn
\sigma_E/(\hbar\omega_L)^2 &=&  
2\left(1-\lambda_c\right)\left(1-\lambda\right) \left|{\cal L}_c\right| \left[ \left|{\cal L}_i\right| 
 -  \sqrt{\left|{\cal L}_i\right| \left(1+\left|{\cal L}_i\right|\right)}\cos(2w) \right]
\nonumber \\ && 
+ 2\left|{\cal L}_c\right| \left(1-\lambda_c\right) + 4\left|{\cal L}_i\right| \left(1+\left|{\cal L}_i\right|\right)
\left(1-\lambda\right),
\eeqnn
where $w=\lambda(v-u/2)$.
We see that results depend on the product $\lambda\lambda_c =\pm 1$, which is positive in the case of ``co-rotation''
of the packet center and ellipse axes and negative for ``anti-rotating'' packets.
The phase difference $w$ does not influence the angular momentum variance (as well as its mean value) in the 
``anti-rotating'' case:
\[
\widetilde\sigma_L = \left|{\cal L}_c\right| + 2\left|{\cal L}_i\right| \left(1+\left|{\cal L}_i\right|\right),  \quad
\lambda\lambda_c=-1.
\]
But this phase is important in the case of ``co-rotation'' (we assume that ${\cal L}_i>0$):
\[
\widetilde\sigma_L = {\cal L} +{\cal L}_i (1+ 2{\cal L}) - 2{\cal L}_c \sqrt{{\cal L}_i \left(1+{\cal L}_i\right)}\cos(2w),
  \quad
\lambda\lambda_c=+1, \quad {\cal L} ={\cal L}_i + {\cal L}_c.
\]
The energy variance equals zero for all packets whose directions of ``internal'' and ``external'' rotations coincide with the direction of the Larmor rotation: $\lambda=\lambda_c=1$. 
The relative phase $w$ is important if only $\lambda=\lambda_c=-1$ (packets performing ``co-rotation'' in the
direction opposite to the Larmor rotation).

Covariances of coordinates and canonical momenta for the minimum energy Gaussian packets 
were calculated in \cite{Drot}). Using that results, the following expressions for the
covariances 
can be obtained (for $\omega_c >0$):
\be
\left.
\begin{array}{c}
\sigma_{XX} \\
\sigma_{YY}
\end{array}
\right\} 
= \frac{\hbar}{2M\omega_c} \left[ 1 +\left(\left|{\cal L}_i\right| + {\cal L}_i\right)
\left(1 \mp\cos(u)/\rho\right)\right], 
\ee
\be
\left.
\begin{array}{c}
\sigma_{\xi\xi} \\
\sigma_{\eta\eta}
\end{array}
\right\} 
= \frac{\hbar}{2M\omega_c} \left[ 1 +\left(\left|{\cal L}_i\right| - {\cal L}_i\right)
\left(1 \mp\cos(u)/\rho\right)\right]. 
\ee
We see that there is no squeezing in the $\xi-\eta$ pair if ${\cal L}_i >0$, as it must be for the
states with the absolute minimum of the energy. At the same time, the $X-Y$ pair becomes squeezed,
and the degree of squeezing can be arbitrarily large for $\left|{\cal L}_i\right| \gg 1$. For example,
if $\cos(u)=1$, then $\sigma_{XX} \approx \hbar/\left(8M\omega_c{\cal L}_i\right)$ 
and $\sigma_{YY} \approx 2\hbar{\cal L}_i/\left(M\omega_c\right)$
for ${\cal L}_i \gg 1$. 
Note that $\sigma_{XX}\sigma_{YY} \equiv \left[\hbar/\left(2M\omega_c\right)\right]^2$
for any ${\cal L}_i >0$
if $|\cos(u)|=1$,
so that in this case we have the minimal uncertainty state for the $X-Y$ pair with respect to the 
commutation relations (\ref{maincomm}).
For ${\cal L}_i <0$, we have no squeezing in the $X-Y$ pair, whereas an arbitrary squeezing
can be achieved for the $\xi-\eta$ pair, if $\left|{\cal L}_i\right| \gg 1$.

Further studies on rotational Gaussian packets were performed in \cite{Drot-mix,Goussev17}.

\section{Non-Gaussian states}
\label{sec-nonG}

\subsection{``Partially displaced'' states}

Coherent states $|\alpha,\beta\rangle$  (\ref{alnm}) are special superpositions of {\em all\/} stationary states 
$|n,m\rangle$. Taking specific sums over the single quantum number $n$ or $m$ one can construct various 
``partially coherent'' states. Malkin and Man'ko \cite{MM69} have constructed two such families of states.
The states with a well-defined energy and the Poissonian distribution over the quantum number $m$ have the form
\be
|n,\beta\rangle = \exp \left(-|\beta|^{2}/{2} \right)
\sum_{m=0}^{\infty}\frac{ \beta^m }{\sqrt{m!}}|n,m\rangle,
\label{nbeta}
\ee
\be
\langle x,y|n,\beta\rangle = \sqrt{\frac{M \omega_c}{2 \pi \hbar n!}}\,i^n
\left(\sqrt2 \zeta^* -\beta\right)^n
	\exp \left(- \zeta \zeta^{*}+ \sqrt{2} \beta \zeta
        -|\beta|^{2}/{2} \right).
\label{MM-nbeta}				
\end{equation}
Such states were considered also in \cite{Loyola89} in order to elucidate the infinite
degeneracy of the energy levels in the case of a uniform magnetic field.

Another family of ``partially coherent'' states considered in \cite{MM69} is
\be
|\alpha,m\rangle = \exp \left(-|\alpha|^{2}/{2} \right)
\sum_{n=0}^{\infty}\frac{ \alpha^n }{\sqrt{n!}}|n,m\rangle,
\label{alm}
\ee
\be
\langle x,y|\alpha,m\rangle = \sqrt{\frac{M \omega_c}{2 \pi \hbar m!}}
\left(\sqrt2 \zeta -i\alpha\right)^m
	\exp \left(- \zeta \zeta^{*}+ i\sqrt{2} \alpha \zeta^*
        -|\alpha|^{2}/{2} \right).
\label{MM-alm}				
\end{equation}
Obviously 
\[
|\alpha,\beta\rangle = \exp\left(-|\alpha|^2 /2\right)
\sum_{m=0}^{\infty}\frac{\alpha^n }{\sqrt{n! }}|n,\beta\rangle
= \exp\left(-|\beta|^2 /2\right)
\sum_{m=0}^{\infty}\frac{ \beta^m}{\sqrt{ m!}}|\alpha,m\rangle.
\]
Similar displaced Landau states were considered later, e.g., in Refs. 
\cite{Ali05,Ferrari90,Mouayn04,Yang07,Rhimi08,Abreu15}.

\subsection{Coherent states with a fixed angular momentum}

The states $|\alpha,\beta\rangle$, $|n,\beta \rangle$ and $|\alpha,m\rangle$ do not possess
a definite value of the angular momentum $L$. However, taking superpositions of the states 
$|n,m\rangle $ (\ref{nmstates}) with a fixed value $l=m-n$ one can construct various families 
of coherent states with a definite angular momentum. 
An explicit example was given in \cite{Fan99}:
\be
|z,l\rangle = {\cal N}\sum_{m=\mbox{max}(0,l)}^{\infty} \frac{z^m}{\sqrt{(m-l)! m!}}
|m-l,m\rangle, \qquad
|{\cal N}|^{-2} = |z|^{-|l|}I_{|l|}(2|z|),
\label{Fan-l}
\ee
where $I_q(x)$ is the modified Bessel function of the first kind.
(The sign of $l$ in the above formula is opposite to that in \cite{Fan99},
due to the different choice of the electric charge sign.)

Since the state $|z,l\rangle$ is an eigenstate of the operator $\hat{L}$ (\ref{HLab}), it 
satisfies the equation
\be
\left(\hat{b}^{\dagger}\hat{b} -\hat{a}^{\dagger}\hat{a}\right) |z,l\rangle = l |z,l\rangle.
\label{lzl}
\ee
But it is easy to see that, in addition, the state $|z,l\rangle$ is an eigenstate of 
operator $\hat{a}\hat{b}$:
\be
\hat{a}\hat{b} |z,l\rangle = z |z,l\rangle.
\label{abz}
\ee
Actually, the states defined by the equalities (\ref{lzl}) and (\ref{abz}) were introduced 
earlier in \cite{BhaBha}, where the operator 
$\hat{b}^{\dagger}\hat{b} -\hat{a}^{\dagger}\hat{a}$ was interpreted as 
the ``charge operator''. Therefore the state (\ref{Fan-l}) was named there ``charged coherent state''.
It can be obtained from the coherent state (\ref{alnm}) by means of the integration \cite{BhaBha}:
\be
|z,l\rangle = \frac{z^{l/2}}{2\pi}e^{|z|}{\cal N}\int_{0}^{2\pi} d\vf e^{-il\vf}
|\sqrt{z} e^{-i\vf},  \sqrt{z} e^{i\vf}\rangle.
\label{zl-int-ab}
\ee
Then using Eq. (\ref{MMcoh}) we obtain the following wave function in the coordinate space:
\be
\langle x,y|z,l\rangle = \sqrt{\frac{M \omega_c}{2 \pi \hbar}} {\cal N}
\left(iz\zeta/\zeta^*\right)^{l/2} J_l \left(2|\zeta|\sqrt{2z}\,e^{-i\pi/4} \right)
\exp\left(-|\zeta|^2 -iz \right),
\label{psizl}
\ee
where $J_l(x)$ is the usual Bessel function.

In view of relation (\ref{abz}), 
the states $|z,l\rangle$ can be considered as some kind of two-dimensional generalizations 
of the {\em Barut--Girardello coherent states\/} \cite{BG}.
The explicit constructions can be found in Refs. \cite{Fakhri04,Aremua15}.

\subsection{$su(1,1)$ and $su(2)$ coherent states}

Using products and squares of the linear annihilation operators $\hat{a}$ and 
$\hat{b}$ (\ref{9-10}), together with their creation partners,
 one can construct various sets of new operators, satisfying the
commutation relations between the generators of the $su(1,1)$ or $su(2)$ algebras,
$\hat{K}_{\pm}$ and $\hat{K}_0$. 
For example, the $su(2)$ case corresponds to the choice \cite{Dehghani13}
\be
\hat{K}_{-} =\hat{a}^{\dagger}\hat{b}, \quad \hat{K}_{+} =\hat{b}^{\dagger}\hat{a},
\quad \hat{K}_{0}= \left(\hat{b}^{\dagger}\hat{b} - \hat{a}^{\dagger}\hat{a}\right)/2,
\ee
\be
\left[\hat{K}_{+},\hat{K}_{-}\right]=2\hat{K}_{0}, \quad
\left[\hat{K}_{0},\hat{K}_{\pm}\right]= \pm\hat{K}_{\pm}.
\ee
The $su(1,1)$ algebra arises for the choice \cite{Aremua15}
\be
\hat{K}_{-} =\hat{a}\hat{b}, \quad \hat{K}_{+} =\hat{b}^{\dagger}\hat{a}^{\dagger},
\quad \hat{K}_{0}= \left(\hat{a}^{\dagger}\hat{a} + \hat{b}\hat{b}^{\dagger}\right)/2,
\ee
\be
\left[\hat{K}_{+},\hat{K}_{-}\right]= -2\hat{K}_{0}, \quad
\left[\hat{K}_{0},\hat{K}_{\pm}\right]= \pm\hat{K}_{\pm}.
\ee
Then, using the Klauder--Perelomov scheme \cite{Klauder63,Perelomov},
 one can construct different families of
states of the form 
$\exp(\zeta_{+}\hat{K}_{+} +\zeta_{-}\hat{K}_{-} +\zeta_{0}\hat{K}_{0})|f\rangle$,
frequently called as the $su(1,1)$ and $su(2)$ coherent states.
Various explicit examples were studied in detail, e.g., in Refs.
\cite{Dehghani12,Fakhri04,Aremua15,Dehghani13,Novaes03,Fakhri03,Setare09}.

\subsection{Semi-coherent states}

In 1973,
Mathews and Eswaran \cite{MaEs73} introduced the notion of
``semi-coherent states'',
defining them as those states of a harmonic oscillator which
possess {\em time-independent\/}
values of the quadrature variances $\sigma_x$ and $\sigma_p$, different from
the vacuum (or coherent state) values.
The necessary and sufficient condition for such
states is
\begin{equation}
\langle \hat{a}^{2}\rangle =\langle \hat{a}\rangle ^{2},
\label{cond}
\end{equation}
where $\hat{a}=(\hat{x}+i\hat{p})/\sqrt2$
is the usual bosonic annihilation operator
(in the units with $\hbar=M=\omega=1$).
The condition (\ref{cond})
is obviously satisfied for the usual coherent states
$|\alpha\rangle$, as soon as $\hat{a}|\alpha\rangle=\alpha|\alpha\rangle$.
Another trivial example is the Fock state $|n\rangle$, for which
$\langle n|\hat{a}^{2}|n\rangle =\langle n|\hat{a}|n\rangle =0$.
A nontrivial example, given in \cite{MaEs73}, is a
normalized superposition of two coherent states of the form
\be
|\alpha\perp\beta\rangle =\frac{|\alpha\rangle -
|\beta\rangle \langle \beta|\alpha\rangle}
{\left(1-|\langle \beta|\alpha\rangle|^2\right)^{1/2}}.
\label{semicoh}
\ee
The notation $|\alpha\perp\beta\rangle$
emphasizes that the state (\ref{semicoh})
is orthogonal to the state $|\beta\rangle$:
$
\langle\beta|\alpha\rangle =0$.
Therefore, the Mathews--Eswaran state 
$|\alpha\perp\beta\rangle$
can be considered \cite{MaEs73} as an orthogonal projection of
the coherent state $|\alpha\rangle$ on another
coherent state $|\beta\rangle$.
The statistical properties of the state (\ref{semicoh}) were studied in detail in Ref.
\cite{DoRe06}.

The two-dimensional generalizations of semi-coherent states (\ref{semicoh}) of the form
$|(\alpha\beta)\perp(\alpha^{\prime}\beta^{\prime})\rangle$ 
(where $|(\alpha\beta)\rangle$ are the Malkin--Man'ko coherent states)
were introduced in \cite{Dehghani15}.
Taking some arbitrarily chosen values of parameters 
$\alpha^{\prime}$ and $\beta^{\prime}$ (e.g., $\alpha^{\prime}=0.1$ and $\beta^{\prime}=0.05$),
the authors have shown that the new states possess the sub-Poissonian statistics and squeezing 
(with respect to the {\em canoniacal\/} momenta) for some values of parameters $\alpha$ and $\beta$.
However, they did not study the squeezing properties with respect to the geometrical pairs
$(\xi,\eta)$ and $(X,Y)$. Therefore, further studies of states
$|(\alpha\beta)\perp(\alpha^{\prime}\beta^{\prime})\rangle$ would be interesting, especially the
search for the most interesting combinations of four complex parameters 
$\alpha,\beta,\alpha^{\prime},\beta^{\prime}$.

\subsection{Nonlinear coherent states}

The general concept of ``nonlinear coherent states'' (NLCS) was introduced 
(for a single degree of freedom) in Refs. 
\cite{Matos96,ManMar97}, although various special cases of such states have been known 
much earlier under other names (see reviews \cite{DR1,Sivakumar00} for details).
 These states were defined
as eigenstates of the product of the boson annihilation operator $\hat{a}$
and some function $f (\hat{n})$ of the number operator $\hat{n} = \hat{a}^{\dagger} \hat{a}$:
\be
 \hat{a} f (\hat{n}) |\alpha, f\rangle = \alpha|\alpha, f \rangle. 
\label{NLCS}
\ee
The decomposition of the state $|\alpha, f \rangle$ over the Fock states reads as \cite{ManMar97}
\be
|\alpha, f \rangle = {\cal N}\sum_{n=0}^{\infty}\frac{\alpha^n }{\sqrt{n! [f(n)]!}}
|n\rangle, \qquad [f(n)]! \equiv f(0) f(1) \cdots f(n),
\label{NLCS-n}
\ee
where ${\cal N}$ is the normalization factor. Therefore the NLCS are close to the Gazeau--Klauder
coherent states introduced in \cite{GK99}.

Two-dimensional NLCS for a charged particles moving in a uniform magnetic field were 
introduced by Kowalski and Rembieli\'nski
\cite{Kowalski05}. These states 
were defined according to the relations
\be
\hat{b}|\zeta,\beta\rangle =\beta |\zeta,\beta\rangle, \qquad
\exp\left(\hat{a}^{\dagger}\hat{a}\right)\hat{a}|\zeta,\beta\rangle =\zeta |\zeta,\beta\rangle,
\label{expfn}
\ee
so that their decomposition over the states $|n,m\rangle$ (\ref{nmstates}) reads as
\be
|\zeta,\beta\rangle = {\cal N}\sum_{n,m=0}^{\infty}\frac{\zeta^n \beta^m}{\sqrt{n! m!}}
\exp\left[-\frac12(n-1/2)^2\right] |n,m\rangle  .
\label{coh-Kow}
\ee
 It was shown in \cite{Kowalski05} that,
according to some criteria, the states (\ref{coh-Kow}) can be better approximations 
of the phase space than the Malkin--Man'ko coherent states (\ref{alnm}).
The comparison between these two families of coherent states was made also in Ref. \cite{Herrera08}.
The generalization  to the case, where the operator
$\exp(\hat{n})$ in  (\ref{expfn})
is replaced by $\exp(\lambda\hat{n})$
with an arbitrary parameter $\lambda \ge 0$, was studied in Ref. \cite{Gazeau09}.

Other kinds of NLCS were constructed in papers \cite{Lev97,Lev02}.
Their authors considered the Klein--Gordon equation in the Feshbach--Villars representation
\cite{Feshbach58}.
Eigenstates of the {\em even part\/} of the annihilation operator $\hat{a}$, describing the relative
motion in the plane perpendicular to the magnetic field, were found in the form (\ref{NLCS})
with the nonlinear function 
\[
f(n) = \frac{E(n-1) +E(n)}{2\sqrt{E(n-1) E(n)}}, \qquad 
E(n)= \sqrt{1 +(2n+1)\hbar\omega_c/(Mc^2)}
\]
(the existence of the second quantum number $m$ was not taken into account).

One more example is the ``angular momentum-phase coherent state'' \cite{Fan01}, i.e., an
eigenstate of the operator 
$
\hat{A} =\sqrt{\hat{a}^{\dagger}\hat{a} -\hat{b}^{\dagger}\hat{b}} \,
\sqrt{\left(\hat{b} +i \hat{a}^{\dagger}\right)\left(\hat{b}^{\dagger} -i \hat{a}\right)^{-1}}$,
with operators $\hat{a}$ and $\hat{b}$ defined in Eq. (\ref{9-10}) (the circular gauge of the
vector potential was assumed).
This operator can be interpreted (in dimensionless variables)
as $\sqrt{\hat{L}_z +1} e^{i\varphi}$, where $\varphi$ is the polar angle in the $xy$ plane.

\subsubsection{``Photon-added states''}

One of many subfamilies of the NLCS contains the so called ``photon-added states''
$|\alpha,q\rangle$, introduced 
(for the 1D harmonic oscillator) by Agarwal and Tara \cite{AgTara91}:
\be
|\alpha,q\rangle = \frac{\hat{a}^{\dagger q}|\alpha\rangle}
{\sqrt{\langle \alpha|\hat{a}^q\hat{a}^{\dagger q}|\alpha\rangle}},
\ee
where $q$ is a non-negative integer. It was shown in \cite{Sivakumar00} that
these states obey the eigenvalue equation 
$f(\hat{n},q)\hat{a}|\alpha,q\rangle = \alpha|\alpha,q\rangle $ with
$f(\hat{n},q) = 1 - q(1+\hat{n})^{-1}$.
The two-dimensional generalization $|\alpha,\beta;q\rangle = {\cal N}\hat{a}^{\dagger q}|\alpha,\beta\rangle$,
where $|\alpha,\beta\rangle$ is the Malkin--Man'ko coherent state, was studied in Ref. \cite{Mojaveri15}.

\subsection{Coherent states for inhomogeneous magnetic fields}
\label{sec-inhom}

Coherent states for the combination of homogeneous and Aharonov--Bohm magnetic fields,
\[
H_z =H_0 +\Phi \delta(x)\delta(y), \qquad A_x = -y\left(\frac{\Phi}{2\pi r^2} +\frac{H}{2} \right),
\quad A_y = x\left(\frac{\Phi}{2\pi r^2} +\frac{H}{2} \right),
\]
were studied in Refs. \cite{Skarzh-176,Bagrov10,Bagrov11,Bagrov12}.
Trajectory-coherent states for this geometry were considered in \cite{Belov93}.
The nonuniform magnetic field $B_z= -\beta/x^2$ was considered in \cite{Setare09}.
The Morse-like (exponentially decaying) inhomogeneous magnetic fields were considered in 
\cite{Fakhri10,Mojaveri13-Morse}.

\subsection{Supersymmetric coherent states, non-commutative planes and \\ non-Euclidean geometries}

Supersymmetric coherent states for a charged particle in a uniform constant magnetic field were studied in
\cite{Beckers88,Fatyga91,Fakhri02,Twareque08}. These states take into account the spin degrees of freedom.
The case of time-dependent uniform magnetic field was considered in \cite{Kost93}.
Generalizations to the case of motion on a non-commutative plane were considered in
\cite{Fiore11,Baldiotti09,Liang10}.
Non-Euclidean geometries were discussed in
\cite{Mouayn05,Mouayn05a,Hall12,Kurochkin16,Salazar16}.

\section*{Acknowledgments}

I thank Prof. J.-P. Antoine and Prof. J.-P. Gazeau for inviting me to the 
conference ``Coherent states and their applications: A contemporary panorama'',
and for the patience, waiting for this contribution.
I am grateful to CIRM--Marseille for covering my local expenses.
A partial support of the Brazilian agency  CNPq is acknowledged.


\end{document}